\documentclass[manuscript]{aastex1}
\usepackage{natbib}
\pdfoutput=1

\begin{document}

\title{A Search for Point Sources of EeV Neutrons}

\author{The Pierre Auger Collaboration\altaffilmark{1}} 


\altaffiltext{1}{www.auger.org.ar; www.auger.org}

\altaffiltext{1}{Pierre Auger Collaboration, Av. San Mart\'{\i}n Norte 306, 5613 Malarg\"ue, Mendoza, Argentina}

\begin{abstract}
A thorough search of the sky exposed at the Pierre Auger Cosmic Ray
Observatory reveals no statistically significant excess of events
in any small solid angle that would be indicative of a flux
of neutral particles from a discrete source.  The search covers from
$-90^{\circ}$ to $+15^\circ$ in declination using four different
energy ranges above 1 EeV ($10^{18}$ eV).  The method used in this
search is more sensitive to neutrons than to photons.  The upper limit
on a neutron flux is derived for a dense grid of directions for each
of the four energy ranges.  These results constrain scenarios for the
production of ultra-high energy cosmic rays in the Galaxy.
\end{abstract}


\linespread{0}
\par\noindent
\begin{center}{\bf The Pierre Auger Collaboration}\\
\end{center}
\begin{small}
P.~Abreu$^{63}$, 
M.~Aglietta$^{51}$, 
M.~Ahlers$^{94}$, 
E.J.~Ahn$^{81}$, 
I.F.M.~Albuquerque$^{15}$, 
D.~Allard$^{29}$, 
I.~Allekotte$^{1}$, 
J.~Allen$^{85}$, 
P.~Allison$^{87}$, 
A.~Almela$^{11,\: 7}$
% ...

\par\noindent
$^{1}$ Centro At\'{o}mico Bariloche and Instituto Balseiro (CNEA-UNCuyo-CONICET), San 
Carlos de Bariloche, 
Argentina \\
% ...
$^{7}$ Instituto de Tecnolog\'{\i}as en Detecci\'{o}n y Astropart\'{\i}culas (CNEA, CONICET, UNSAM), 
Buenos Aires, 
Argentina \\
% ...
$^{11}$ Universidad Tecnol\'{o}gica Nacional - Facultad Regional Buenos Aires, Buenos Aires,
Argentina \\
% ...
$^{15}$ Universidade de S\~{a}o Paulo, Instituto de F\'{\i}sica, S\~{a}o Paulo, SP, 
Brazil \\
% ...
$^{29}$ Laboratoire AstroParticule et Cosmologie (APC), Universit\'{e} Paris 7, CNRS-IN2P3, 
Paris, 
France \\
% ...
$^{51}$ Istituto di Fisica dello Spazio Interplanetario (INAF), Universit\`{a} di Torino and 
Sezione INFN, Torino, 
Italy \\
% ...
$^{63}$ LIP and Instituto Superior T\'{e}cnico, Technical University of Lisbon, 
Portugal \\
% ...
$^{81}$ Fermilab, Batavia, IL, 
USA \\
% ...
$^{85}$ New York University, New York, NY, 
USA \\
% ...
$^{87}$ Ohio State University, Columbus, OH, 
USA \\
% ...
$^{94}$ University of Wisconsin, Madison, WI, 
USA \\
% ...
\par\noindent
(\ddag) Deceased \\
(a) at Konan University, Kobe, Japan \\
(b) now at the Universidad Autonoma de Chiapas on leave of absence from Cinvestav \\
(f) now at University of Maryland \\
(h) now at NYU Abu Dhabi \\
(i) now at Universit\'{e} de Lausanne \\



\end{small}

\newpage
\section{Introduction}\label{introduction}

Neutrons travel on straight lines, undeflected by magnetic fields, and
they produce air showers that are indistinguishable from air showers
produced by protons.  A flux of neutrons from a discrete source would
cause an excess of cosmic ray events around the direction
to the source, clustered within the angular resolution of the
observatory.  Since free neutrons undergo beta decay with a mean
lifetime of about 886 seconds at rest \citep{NeutronLifetime}, the mean
travel distance for relativistic neutrons is $9.2\times E$ kpc, where
$E$ is the energy of the neutron in EeV (1 EeV = $10^{18}$ eV).  The
distance from Earth to the Galactic Center is about 8.3 kpc
\citep{GCDist}, and the radius of the Galaxy is approximately 15 kpc.
Sources in part of the Galactic disk, including the Galactic Center,
should be detectable via neutrons above 1 EeV.  Above 2 EeV, the
volume for detectable neutron emitters includes most of the Galaxy.

An important unresolved issue about high energy cosmic rays is the
transition at some energy from cosmic rays produced in the Galaxy to
cosmic rays from extragalactic sources.  The ankle of the energy
spectrum can be explained as a dip caused by $e^\pm$ production in
collisions of predominantly extragalactic protons with CMB photons
\citep{Blumenthal,Berezinsky}.  A time-honored alternative view is that
the ankle of the spectrum near 4 EeV \citep{PLB} is the transition from
a Galactic power-law spectrum to a harder extragalactic power-law
spectrum \citep{Hillas1972}.  If sources in the Galaxy are emitting
protons up to the ankle of the spectrum, they could show themselves
through a flux of EeV neutrons or perhaps a flux of EeV photons.

The signature of a neutron flux is a simple excess of proton-like air
showers from a single celestial direction.  In contrast, special
discrimination techniques should be used to optimize sensitivity to
EeV photons.  A search for EeV photon fluxes will be reported
separately.  The method used in this paper is not sufficiently
sensitive to photon fluxes because muon-poor photon showers produce
less signal in water Cherenkov detectors than proton showers of the
same energy.  This search is optimized for neutron fluxes, and the
upper limits do not apply to photon fluxes.

Concerning the production of neutrons and photons by energetic
protons, in both cases the dominant process is pion-producing
interactions with ambient photons, protons, or nuclei.  Two photons
are produced by the decay of each $\pi^0$.  Neutrons are produced by
charge-exchange interactions in which a $\pi^+$ takes the positive
charge of the proton and a leading neutron emerges with most of the
energy that the proton had.  The production of neutrons and photons
has been studied extensively \citep{Watson,Bossa,Aharonian,Crocker},
especially in relation to evidence from AGASA \citep{Hayashida} and
SUGAR \citep{Bellido} for possible fluxes from directions close to the
galactic center, which were not confirmed using Auger data \citep{GC}.
Because photons acquire only a small fraction of the proton energy,
the production of neutrons exceeds the hadronic production of photons
of the same energy provided the accelerated proton spectrum falls
approximately like $1/E^2$ or more steeply with energy.

Based on the energy flux of gamma rays measured at TeV energies, known
Galactic sources could plausibly be producing neutron fluxes which
would be detectable by the Pierre Auger Observatory.  The energy flux
in TeV gamma rays exceeds 1 eV/cm$^2$/s at Earth for some Galactic
sources \citep{Hinton,Grillo}.  Suppose those gamma rays arise from
the decay of $\pi^0$ mesons produced by interactions of protons
accelerated at the source.  A source with a $1/E^2$ differential
energy spectrum puts equal energy in each decade, and most models for
cosmic ray sources favor a power-law spectrum of approximately
$1/E^2$.  If there are such sources in the Galaxy producing EeV
photons as well as TeV gamma rays, then the energy flux of EeV photons
should also exceed 1 eV/cm$^2$/s at Earth.  For sources closer than
the neutron attenuation length ($9.2\times E$ kpc), the energy flux of
neutrons could be even higher than the flux of photons above energy
$E$ because the production rate of neutrons at the source should
exceed the production rate of photons of the same energy, as noted in
the previous paragraph.  Using three different energy thresholds, the
results reported here show that there are no EeV sources of neutrons
that bright in the southern sky that is exposed to the Auger
Observatory.

Following a description of the data set (Section \ref{data_set}), 
the method used (Section \ref{Method}), and the uncertainties
(Section \ref{Uncertainties}), the results of this blind
search for a pointlike neutron flux are reported in Section
\ref{blind_search}.  Differential and integral plots of the Li-Ma
significances \citep{Li-Ma} are presented, and upper limits are
plotted on maps of the exposed sky.  These results are then summarized
and discussed in Section \ref{discussion}.  A preliminary version of
this study, using a slightly smaller data set, has been reported
\citep{icrc2011}.

\section{The data set}\label{data_set}

The Pierre Auger Observatory \citep{NIM_EA} is located in Malarg\"{u}e,
Argentina, at latitude 35.2 S, longitude 69.5 W and mean altitude 1400
meters above sea level (870 g cm$^{-2}$ atmospheric depth).  The
surface detector array consists of 1660 water-Cherenkov stations
covering an area of about 3000 km$^2$ on a triangular grid with 1.5 km
spacing, allowing secondary muons, electrons, and photons to be
sampled at ground level with a duty cycle of nearly 100\%.

The data set analyzed here consists of events recorded by the surface
detector (SD) from 1 January 2004 to 30 September 2011. During this
time, the size of the Observatory increased from 154 to 1660 SD
stations.  Events used in this analysis have zenith angles less than
$60^\circ$.  Moreover, an event is accepted only if all six nearest
neighbors of the station with the highest signal were operational at
the time of the event.  This is the standard geometrical aperture cut
that ensures good event reconstruction \citep{Acceptance}.  Periods of
array instability have been omitted from the data set.  The total
exposure of the array with these cuts is 24,880 km$^2$ sr yr
for the period of time analyzed here, yielding 429,138 events with
$E\geq 1$ EeV.

The arrival direction of a cosmic ray is determined from a fit to the
arrival times of the shower front at the SD stations. The precision
achieved in the arrival direction depends on the clock resolution of
each detector and on the fluctuations in the time of arrival of the
first particle \citep{AngRes}. The angular resolution is defined as the
radius of the circular solid angle that would include 68\% of the
reconstructed events that arrive from a fixed direction.  The angular
resolution depends on energy as stated in section \ref{AR}.

The energy of a given air shower is measured by fitting for the ground
signal $S(1000)$ that a station would have measured at 1000 meters
from the core.  This is converted to the energy parameter $S_{38}$,
which is independent of zenith angle.  The energy parameter $S_{38}$
has been calibrated using the quasi-calorimetric air fluorescence
detector.  See \citep{Spectrum_PRL,Pesce} for details about the SD
energy determination.  There is a systematic uncertainty of 22\% in
the absolute energy calibration.  Statistical uncertainty in the SD
energy determination is approximately 15\%.

\section{Method}\label{Method}

\subsection{Energy Cuts}
Four energy ranges are used for the blind search and for the upper
limit analysis: 1 EeV $\leq E <$ 2 EeV (319,818 events), 2 EeV $\leq E
<$ 3 EeV (61,059 events), $E \geq$ 3 EeV (48,261 events), as well as
$E \geq$ 1 EeV.  The first three are independent data sets, while the
final cumulative data set should give maximum sensitivity to a flux
that extends over the entire energy range.  A high-energy range allows
detection of more distant neutron sources in the Galaxy with reduced
cosmic ray background.  A low-energy range favors nearby sources.

\subsection{Target Sizes and Angular Resolutions}\label{AR}
Sensitivity to point sources is optimized by choosing the target size
according to the angular resolution of the SD.  This angular
resolution $\psi$ corresponds to the 68\% containment radius for each
energy. The point spread function is taken to be $p(\theta) =
\frac{\theta}{\sigma^{2}}\exp(-\theta^2/2\sigma^{2})$, where $\theta$
is the angle between the reconstructed direction and the true arrival
direction.  The 68\% containment definition for the angular resolution
$\psi$ means that $\sigma$ can be identified as $\psi/1.51$. With the
choice of a top-hat counting region (selecting events within a hard
cut on angle from the target center), the signal-to-noise ratio is
optimized by top-hat radius $\chi$ given by $\chi = 1.59\sigma =
1.05\psi$.

The angular resolution of the SD has dependence on energy and improves
somewhat at large zenith angles.  Because some declinations are only
viewed at large zenith angles, there is a modest dependence of the
angular resolution $\psi$ on declination as well as energy.  The
median target radius $\chi$ is $1.36^\circ$ for 1-2 EeV, $1.02^\circ$ for 2-3
EeV, $0.69^\circ$ for $E\geq 3$ EeV, and $1.23^\circ$ for $E\geq 1$ EeV.

\subsection{Simulation Data Sets}\label{simulation_data_sets}
To recognize the existence of an excess of events in any
solid angle ``target'', it is necessary to know the number that are
expected in that target without the neutral flux.  Simulation data
sets are used for this.  The expected number of events in a given
target is taken to be the average number found in 10,000 simulated
data sets.  

The simulation data sets are obtained from the actual arrival
directions, for each energy range, by a scrambling procedure that
thoroughly smooths out any small-scale anisotropy.  Each simulation
data set has the same number of arrival directions as the actual data
set.  An arrival direction is produced by randomly sampling a sidereal
time from the set of measured sidereal times, a zenith angle from the
set of measured zenith angles, and an azimuthal angle from a uniform
distribution over $2\pi$ radians.  Each simulation data set should be
equivalent to the actual data aside from statistical fluctuations,
unless astrophysical fluxes have imprinted small-scale anisotropy in
the actual data.

\begin{figure}[ht]
\includegraphics[width=0.9\textwidth]{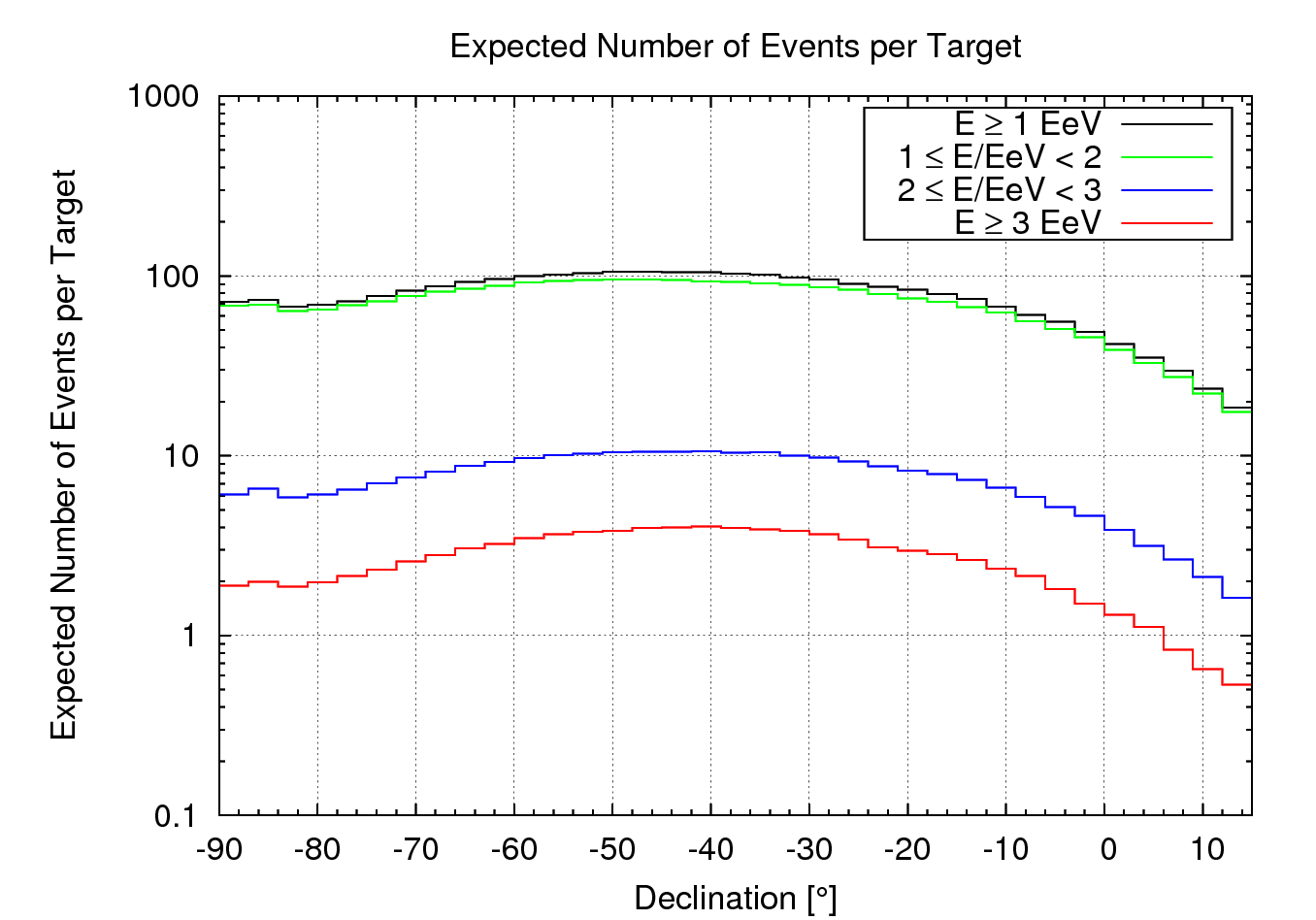} 
\caption{The expected number of events per target for each of the four
  energy ranges, averaged in 3-degree bands of declination.}
\label{Background}
\end{figure}

Figure \ref{Background} shows the expected number of events per target,
averaging over the targets with centers in 3-degree bands of
declination.  For each energy, the expected number depends on
declination partly because of the declination dependence of the target
size, but primarily because the directional exposure varies
with declination.

\subsection{Li-Ma Significance}

The statistical significance $S$ of an excess (or deficit) in a given
target is based on the number of events $n$ observed in
the target, the number $b$ expected in the target from
background cosmic rays, and the Li-Ma parameter $\alpha$:
\begin{equation}
S = \frac{n-b}{|n-b|} \sqrt{2} \left\{ n \ln \left(\frac{n+\alpha n}{b+ \alpha n}\right) +
\frac{b}{\alpha} \ln \left( \frac{b + \alpha b}{b + \alpha n } \right) \right\}^{1/2}.
\end{equation}
This formula is equation 17 of \citep{Li-Ma} using $N_{on}\equiv n$ and
$N_{off}\equiv b/\alpha$.  In gamma ray astronomy, $\alpha$ is the
ratio of time spent observing on-source to the time spent observing an
equivalent off-source solid angle.  For the analysis here, all
off-source regions are used in estimating the background, so
$\alpha$ is taken to be the expected number in the target region
divided by the expected number in the remainder of the sky.

\subsection{Upper Limit Calculation}\label{limit_calc}

There are alternative ways to define the upper limit $s_{UL}$ of
confidence level $CL$ for the {\it expected} signal $s$ when an
observation results in a count $n$ in the presence
of a Poisson background distribution with mean value $b$.  The
definition for $s_{UL}$ adopted here is that of Zech \citep{Zech}:
\begin{equation}
	P(\leq n | b + s_{UL}) = (1 - CL) \times P(\leq n | b)
\end{equation}
where CL is the fractional confidence level (e.g. 95\% confidence
level $\Rightarrow CL=0.95$ and $1-CL=0.05$). This definition avoids
unphysical negative upper limits that occur in the classical
definition when the observed number is a strong downward fluctuation
of the background.  The frequentist
interpretation \citep{Zech} of the above equation is: ``For an
infinitely large number of repeated experiments looking for a signal
with expectation $s_{UL}$ and background with mean $b$, where the
  background contribution is restricted to a value less than or equal
  to $n$, the frequency of observing $n$ or fewer events is
$\alpha$.''  This definition of the upper limit agrees, in this case
of a Poisson process, with the Bayesian upper limit with flat
  prior:
\begin{equation}
\int_0^{s_{UL}} P(n|b+s) ds = (CL)\int_0^{\infty} P(n|b+s)ds.
\end{equation}

This upper limit $s_{UL}$ is for the expected number of events
from the source that would be contained within the top-hat
target region.  For the top-hat radius and assumed point spread
function described in section \ref{AR}, that top-hat region is
expected to encompass 71.8\% of the total signal.  The upper limit
for the total expected number of events from a source in
the direction of the target center is obtained by scaling $s_{UL}$ by
the factor 1/0.718=1.39.

\subsection{Flux Upper Limit}\label{flux-limit}

\begin{figure}[ht]
\includegraphics[width=0.9\textwidth]{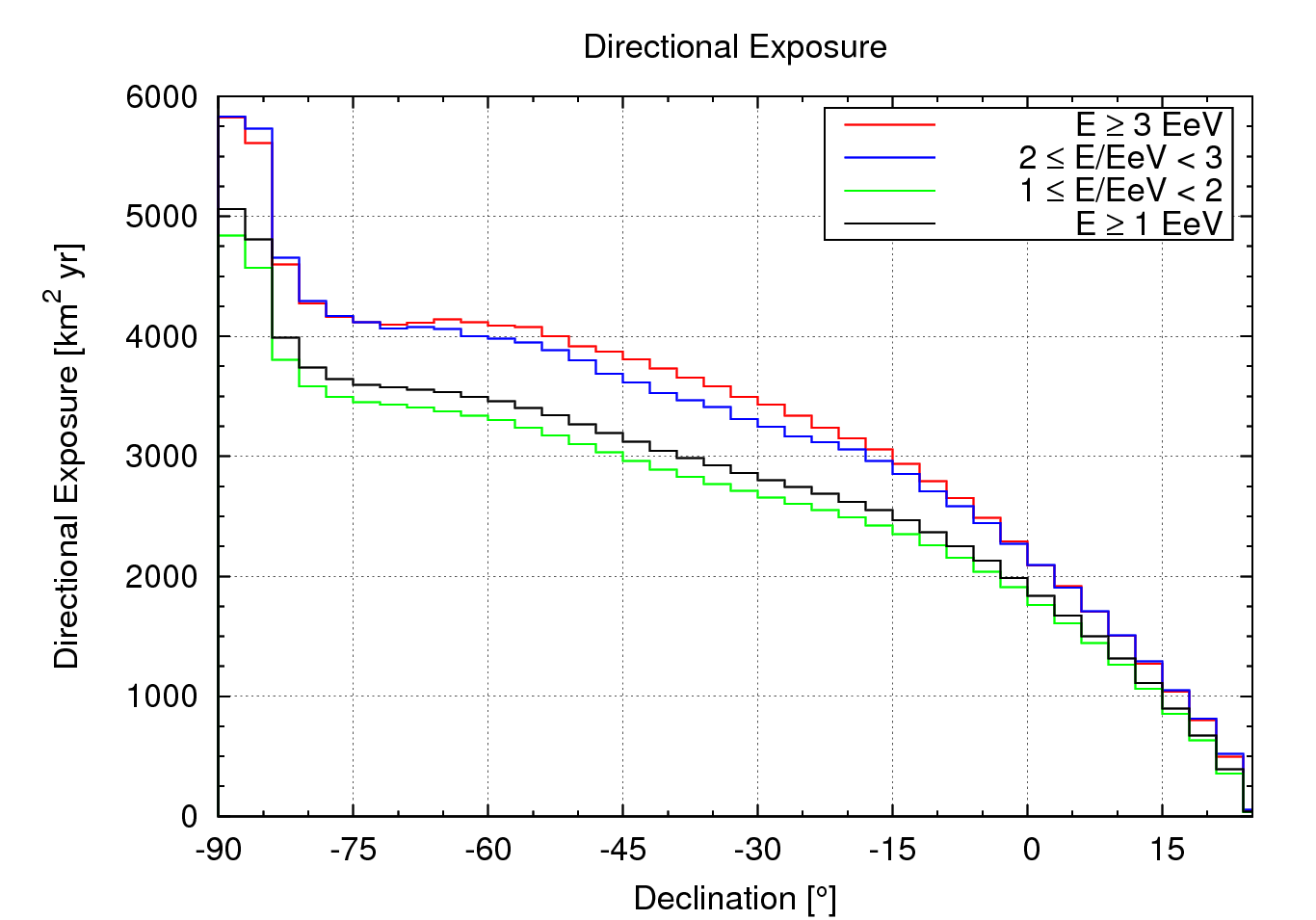} 
\caption{The directional exposure for each of the four
  energy ranges, averaged in 3-degree bands of declination.}
\label{Acceptance}
\end{figure}

The {\it flux upper limit} is the upper limit on the number of events
(as described above) divided by the {\it directional exposure} at the
target center.  The exposure depends on the trigger efficiency: this
is 100\% for energies above 3 EeV \citep{NIM_EA}.  Below that energy,
the efficiency can depend on energy, zenith angle, and primary mass of
the cosmic ray.  The directional exposure for any celestial direction
is given by $\frac{b}{\omega I}$.  Here $b$ is the expected number
(obtained empirically from the average of simulation data sets) in the
target of solid angle $\omega$, and $I$ is the cosmic ray intensity
with units (km$^2$ sr yr)$^{-1}$ calculated by integrating the known
energy spectrum \citep{ICRC_spectrum} over the relevant energy range.
The directional exposure is measured in units of km$^2$yr.  The
dependence of the directional exposure on declination is shown in
Figure \ref{Acceptance} for the four different energy ranges.  (Note
that celestial points within $5^\circ$ of the south pole are constantly
exposed to the array at zenith angles between $50^\circ$ and
$60^\circ$.)  For energy ranges that include energies below 3 EeV, the
empirically derived $b$ includes an implicit efficiency factor for
triggering (generally less than unity) which depends on energy.  The
trigger efficiency depends also on the zenith angle of arrival, causing
a slight dependence of the trigger efficiency on declination
since the distribution of zenith angles varies with declination.

This empirical determination of the directional exposure for an energy
interval implicitly uses the cosmic ray energy spectrum in weighting
the average energy dependence of the directional exposure over that
interval, whereas the (unknown) energy spectrum of a possible neutron
flux would provide the ideal set of weights.  This is not an issue
above 3 EeV or for any narrow energy interval.  Results are reported
for two separate energy bins below 3 EeV partly to reduce this
uncertainty in deriving a flux upper limit from the upper limit on the
number of particles.

The flux upper limit is an upper limit on the {\it time-averaged} flux
based on the cumulative Auger exposure.  Periodic, episodic, or
transient fluxes may exceed these time-averaged limits.  

\subsection{Pixelation and Target Spacing}
The directional exposure of the Auger SD falls rapidly for
declinations close to $+25^\circ$, which is the maximum declination
that can be observed at $-35^\circ$ latitude with zenith angle less
than $60^\circ$.  To avoid excessively large statistical fluctuations,
the search for point sources has been limited to the region where the
directional exposure is greater than 1000 km$^2$yr for a point source,
which means declinations below $+15^\circ$.

HEALPix \citep{Healpix} is used in producing the observed and expected
celestial maps.  The target centers are taken as the central points of
a HEALPix grid with N$_{side}$=128.  The separation of those target
centers is approximately $0.6^{\circ}$.  Since the target diameters
are considerably larger than this separation, they overlap strongly
and are not statistically independent trials.  (This is necessary to
avoid failure to detect a neutron flux simply because it is divided
between two or more adjacent targets.)

For computational efficiency, events are counted and stored in a finer
pixelation with N$_{side}$=512 (approximately 0.015 degree pixel
width).  Instead of being perfect circles, the targets are defined as
the union of the small pixels whose centers lie within the target
radius.  On average, there are 342 small pixels in the targets used
for energy ranges going down to 1 EeV, and there are on average 106
small pixels in the smaller targets that are used for E$>$3 EeV.  Any
data set (actual or simulated) is summarized by the counts of events
in the small pixels.  The number of events in any target is the sum of
the counts in its constituent pixels.

\section{Uncertainties}\label{Uncertainties}

Statistical uncertainty comes from Poisson fluctuations in the number
of events in each target.  The Li-Ma significance and the
flux upper limit at a fixed confidence level are designed to account
appropriately for the Poisson statistical fluctuations.  As such,
statistical error is not a concern.

A systematic error in the angular resolution can affect the overall
normalization of the upper limits.  The mean particle upper limit
scales approximately with the square root of the expected number $b$
in a target, and is therefore proportional to the target radius or the
assumed angular resolution.  The systematic uncertainty in the angular
resolution for each energy bin is approximately 10\%.  Also, a genuine
signal would be underestimated (overestimated) if measured within a
fixed target radius proportional to an assumed angular resolution that
is too small (large).  The fractional error in the top-hat measurement
is 0.714 times the fractional error in angular resolution for small
fractional errors, which means about 7\% uncertainty due to a 10\%
uncertainty in the angular resolution.  The specific shape of the
point spread function assumed in Section \ref{AR} (a Rayleigh
distribution) has little impact on the results.  Any similar
distribution having 68\% containment within angular radius $\psi$
should contain a fraction of the signal flux within the top-hat radius
$1.05\psi$ that is not much different from the 71.8\% Rayleigh
expectation.

In principle, there could be error in the expected background counts
due, for example, to imperfection in the smoothing procedure based on
simulation data sets.  However, those uncertainties are minuscule
compared to Poisson fluctuations in all parts of the sky and for all
energy ranges.  Alternative methods for obtaining the expected counts
give almost identical results.  The uncertainty in expected counts is
small and has a negligible impact on the results.  Similarly, there
could be some systematic uncertainty in the Li-Ma significances
stemming from how the Li-Ma $\alpha$ parameter has been identified in
this context.  Small uncertainty in the background expectations
corresponds to small $\alpha$, and the Li-Ma significances are stable
against changes in $\alpha$ as long as it remains small.  The results here
would be the same for alternative identifications of $\alpha$ that
similarly imply insignificant uncertainty in the background counts.

The results are presented for fixed energy intervals.  If there is a
systematic error in the energy normalization, then the Li-Ma
significance and the flux upper limit for each target pertain to a
different true energy range.  The Auger energy scale
presently has a systematic uncertainty of 22\% \citep{ICRC_spectrum}.
It should be noted that the SD is not fully efficient below 3
EeV, and the trigger inefficiency can introduce a systematic energy
error by favoring upward fluctuating signals in the surface stations.
Events measured in hybrid mode by air fluorescence profiles as well as
by the SD alone indicate that SD energy assignments are systematically
high by about 2\% for 2-3 EeV and close to 7\% for the 1-2 EeV range.

In addition, fluctuations in energy measurements can cause unequal
migration of signal and background events into (and out of) an energy
range, thereby affecting apparent signals and upper limits.  The
effect depends on the exact shape of the arriving neutron signal
spectrum, including its suppression at low energies due to in-flight
neutron decays.  A non-negligible underestimation in an upper limit
could exist if there were no arriving neutrons to spill upward into an
energy bin which does acquire background events by upward fluctuation
in energy measurements.  The errors are not large, however, since the
energy measurement fluctuations are small compared to the size of the
energy ranges, and the background contamination from outside an
energy range does not exceed a few percent.

As explained in Section \ref{flux-limit} and seen in Figure
\ref{Acceptance}, the directional exposure has some dependence on
energy.  Ideally one would use the weighted average for the spectrum
of a hypothetical neutron flux, but the method here uses a weighted
average using the cosmic ray energy spectrum.  For example, a neutron
flux from a distant source might be fully attenuated below 3 EeV, so
the trigger efficiency is 100\% for that flux.  For a measurement
above 1 EeV, however, the directional exposure using the cosmic ray
spectrum might be 20\% lower (see Figure \ref{Acceptance}).  The upper
limit would therefore be (conservatively) too high by about 20\%.
This systematic uncertainty can be reduced by considering separately
the three differential energy ranges where the range of energies is
small for the intervals below 3 EeV.

Some systematic uncertainty exists due to the uncertain cosmic ray
composition.  There is good evidence for a mixed composition which
includes protons throughout the EeV energy decade \citep{Composition},
and neutron showers are reconstructed the same as proton showers.  If
the composition were purely heavy nuclei, the limits here would
pertain to somewhat higher neutron energies than stated.  For a mixed
composition with a substantial proton component, however, energies
reconstructed for neutron showers based on SD data do not differ
systematically more than about 5\% from the background cosmic ray energies
which are calibrated using air fluorescence measurements.  

\section{Results for blind searches}\label{blind_search}

\subsection{Li-Ma Significances}\label{LiMa_results}

\begin{figure}[htbp]
\begin{tabular}{cc}
	\includegraphics[width=0.4\textwidth]{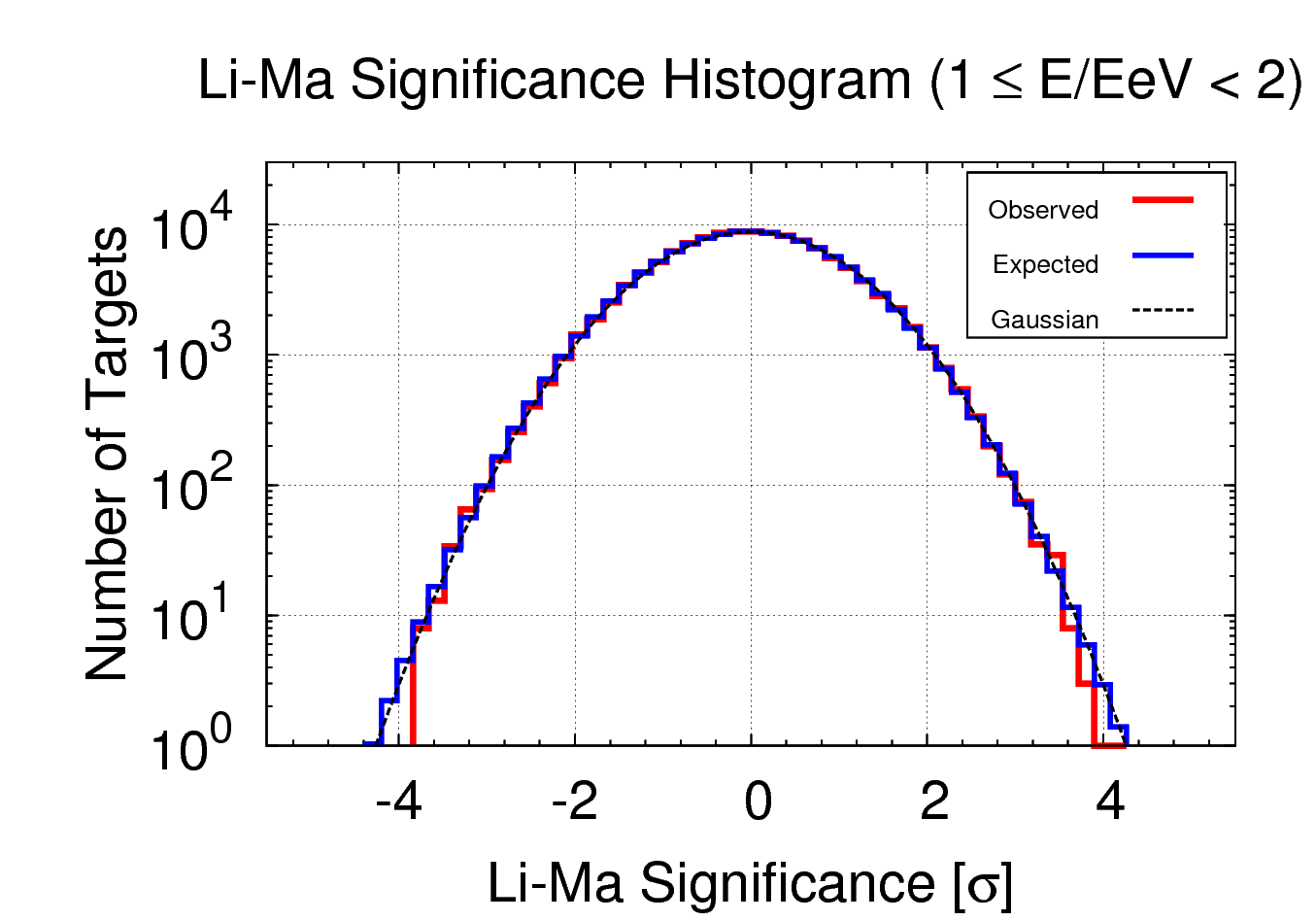} &
	\includegraphics[width=0.4\textwidth]{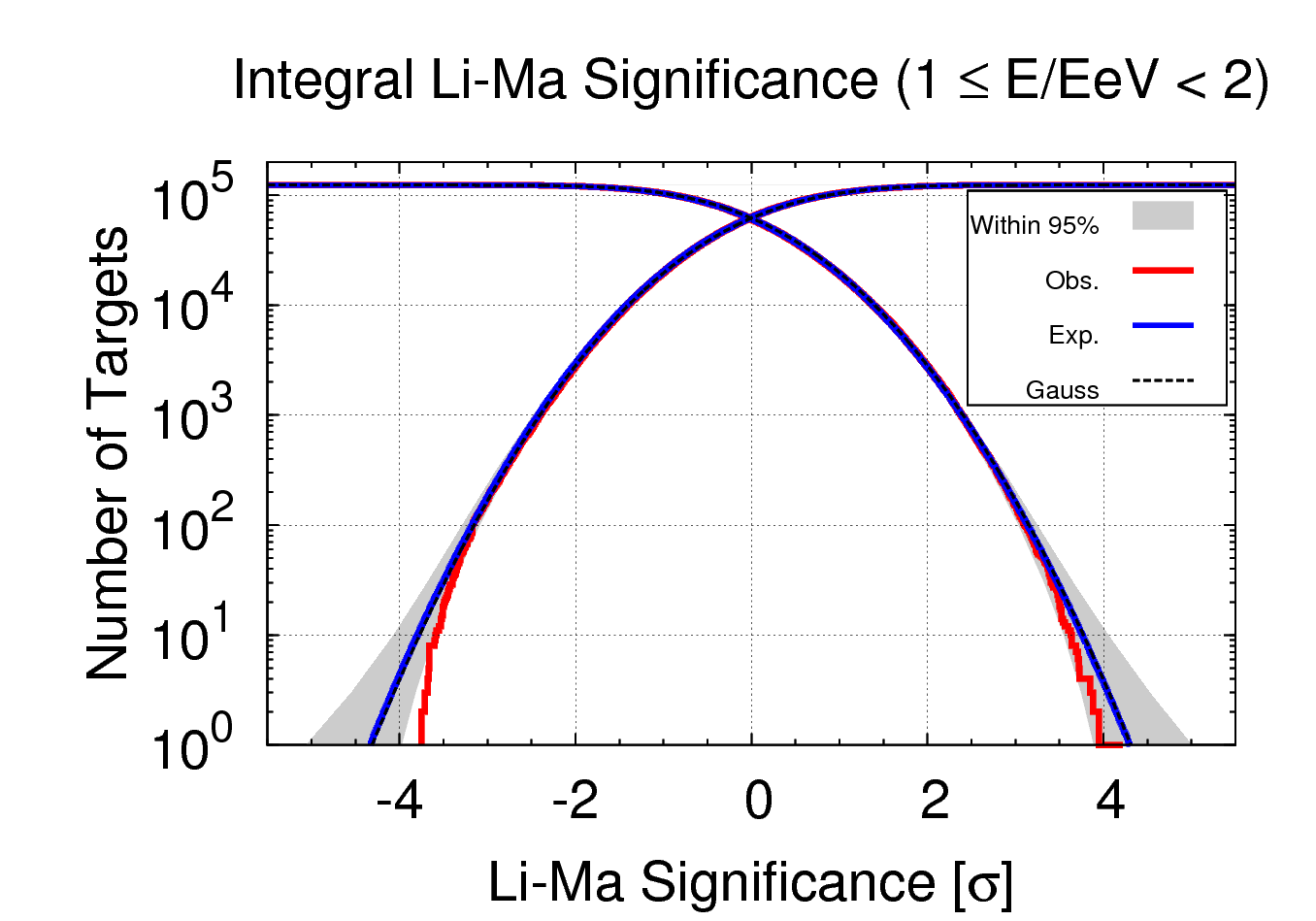} \\
	\includegraphics[width=0.4\textwidth]{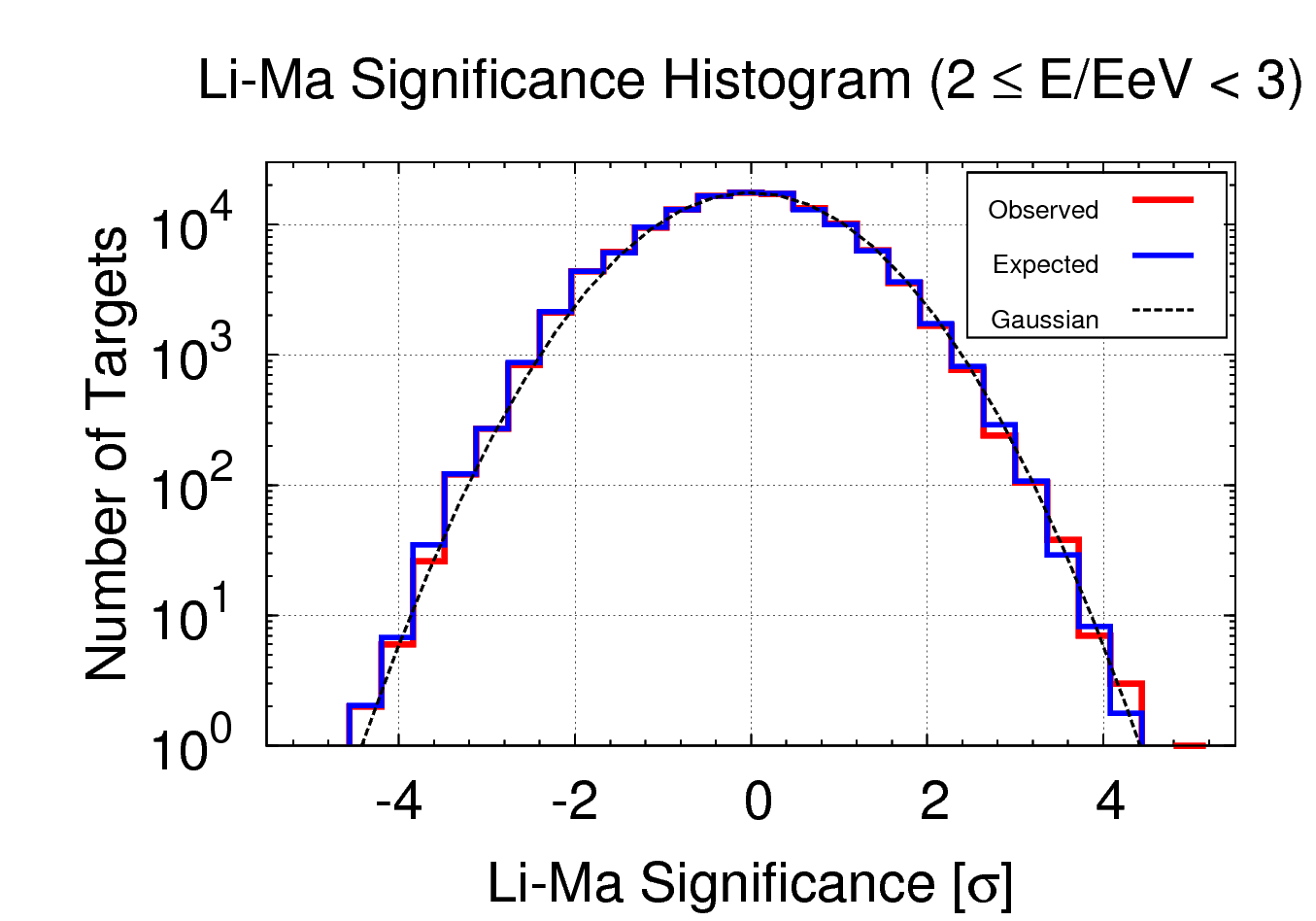} &
	\includegraphics[width=0.4\textwidth]{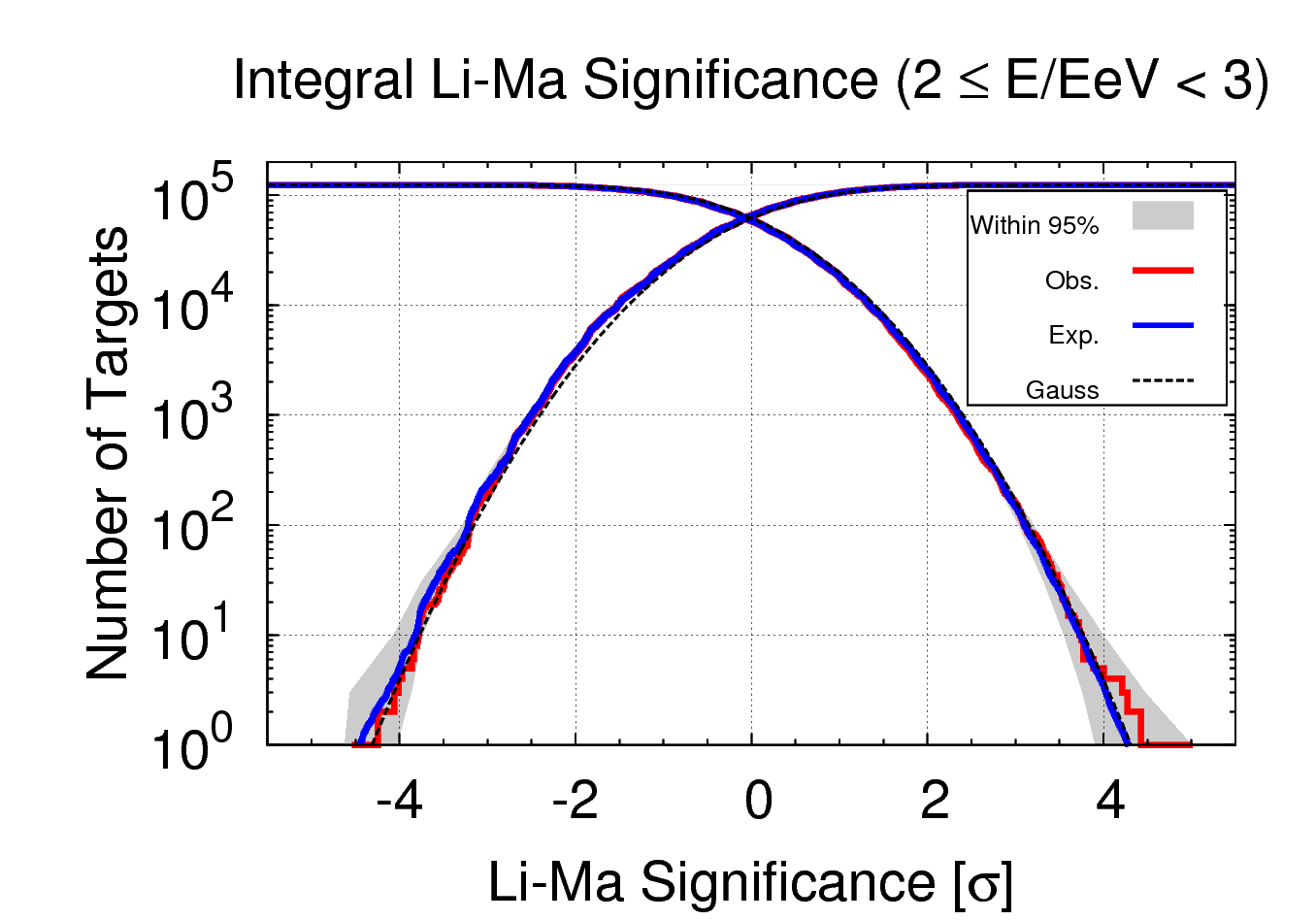} \\
	\includegraphics[width=0.4\textwidth]{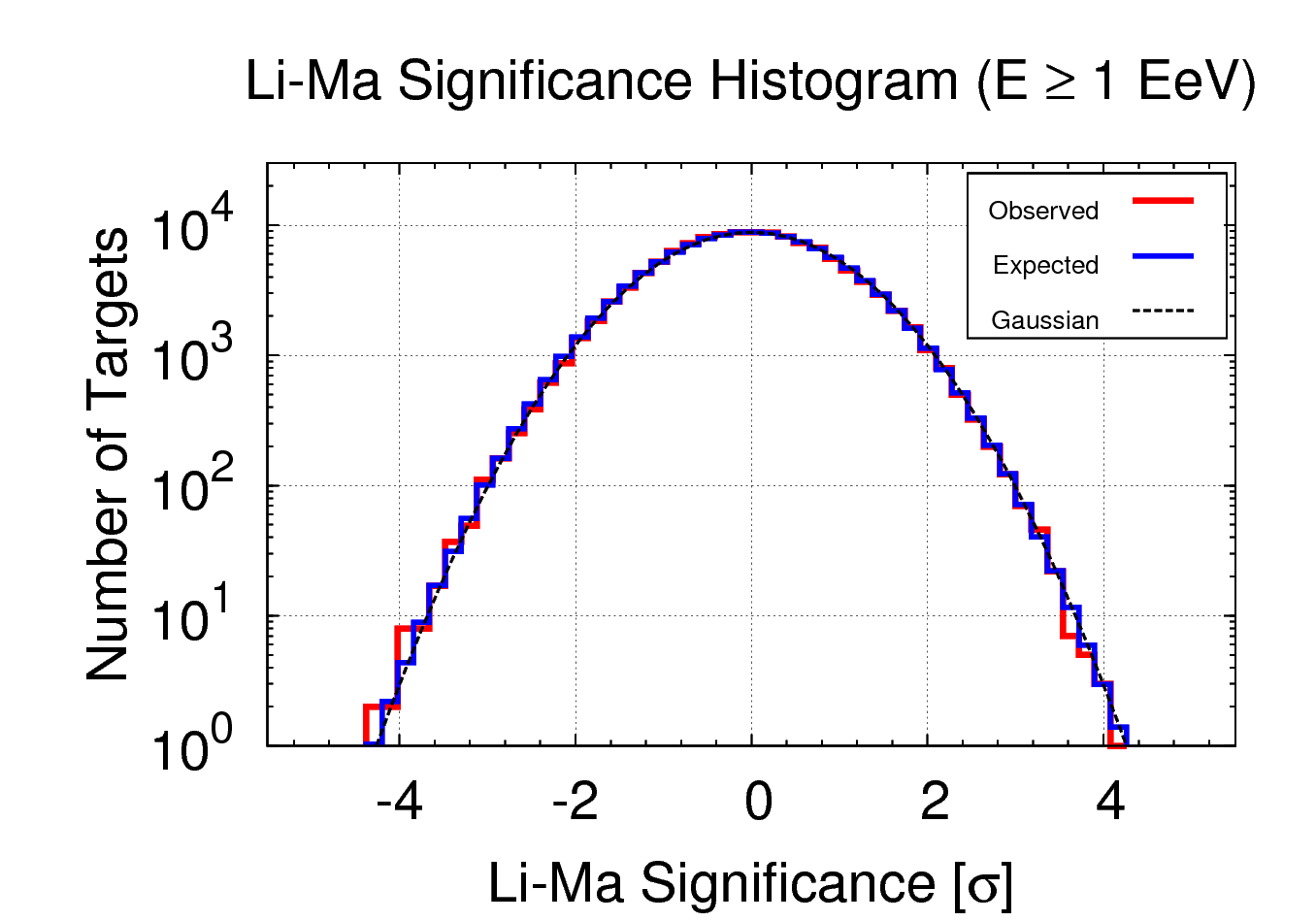} &
	\includegraphics[width=0.4\textwidth]{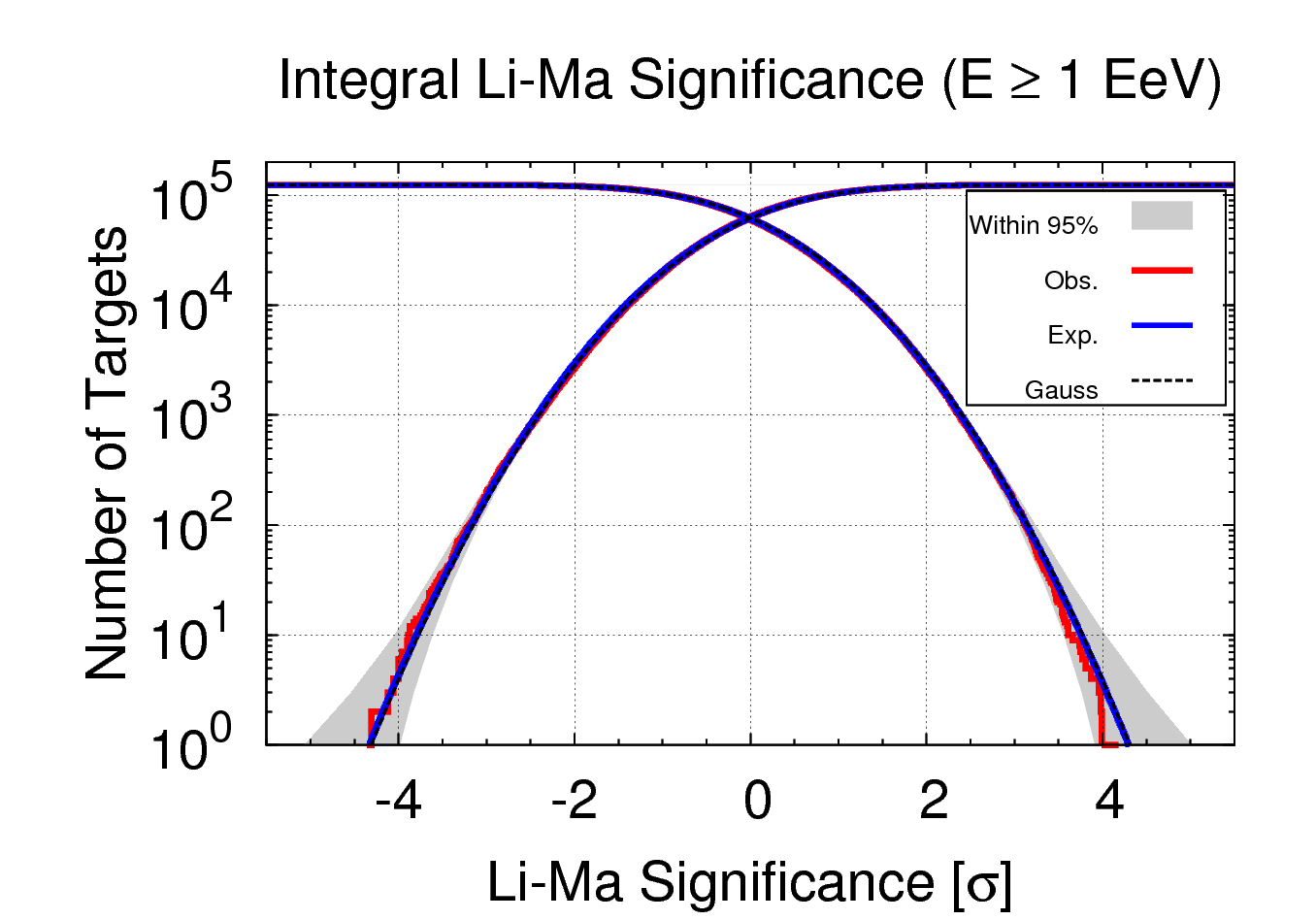} \\
	\includegraphics[width=0.4\textwidth]{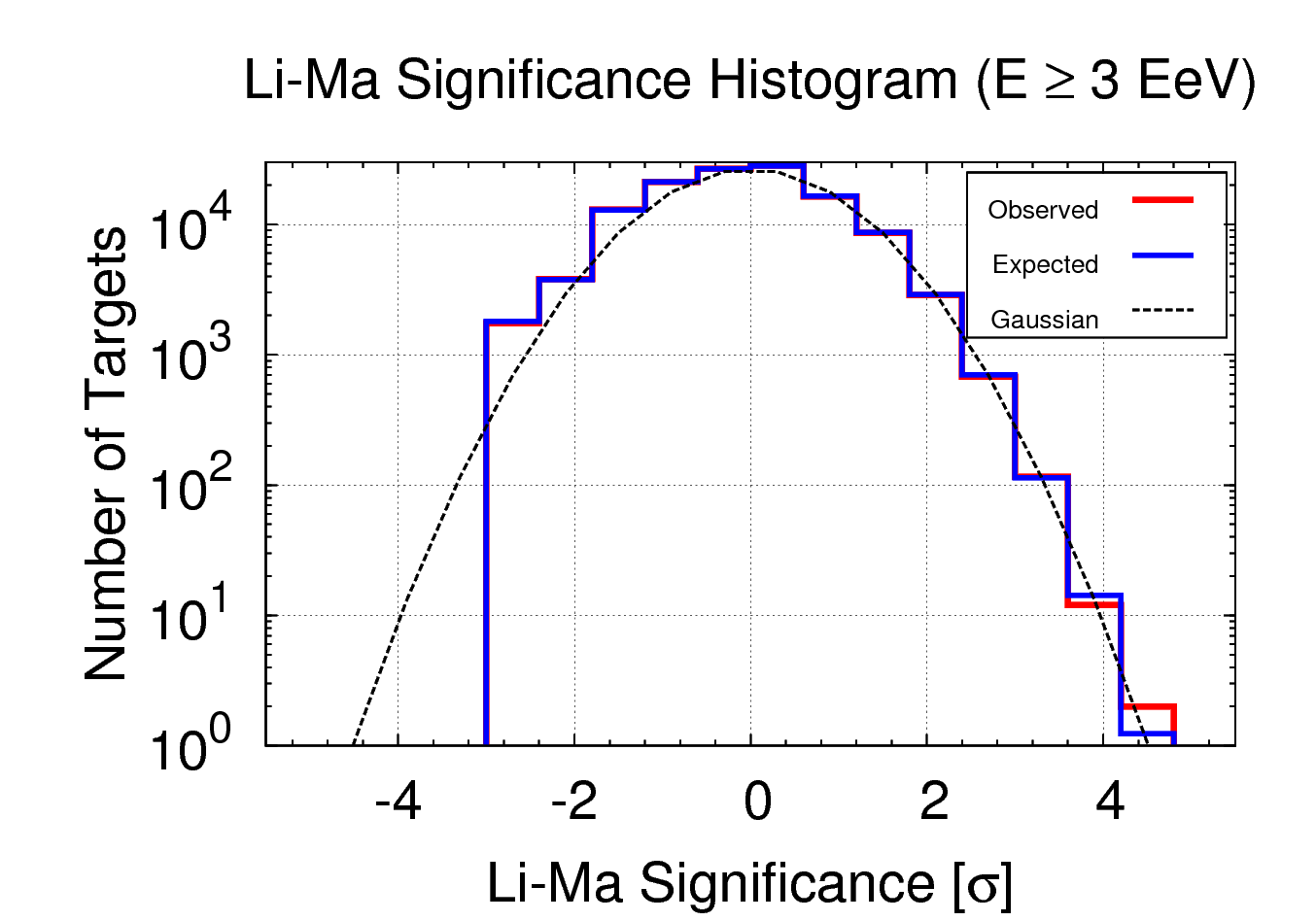} &
	\includegraphics[width=0.4\textwidth]{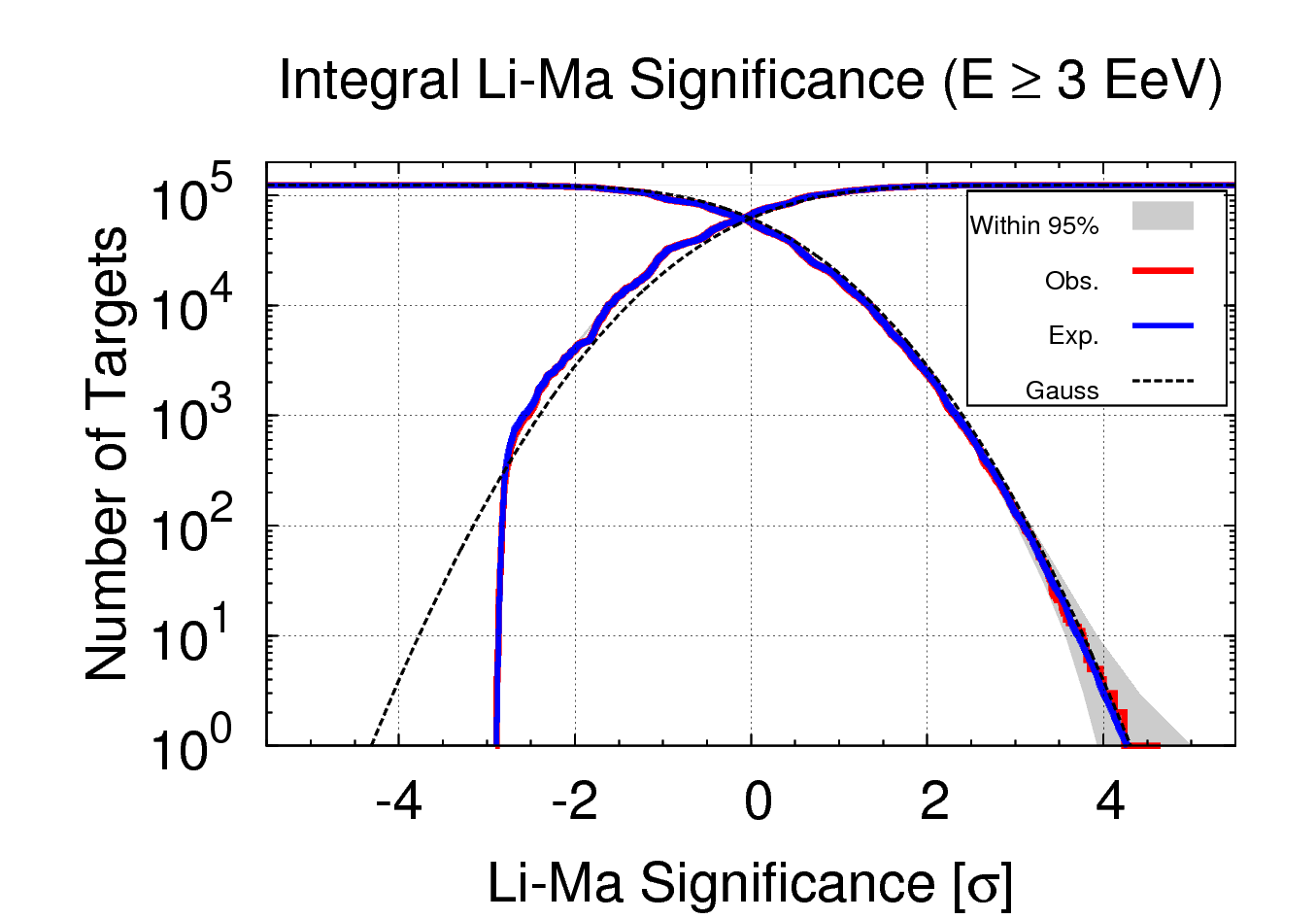} \\
\end{tabular}
\caption{Differential distributions (left) and integral distributions
  (right) of Li-Ma significance for the four energy cuts (1-2, 2-3,
  $\geq 1$, and $\geq 3$ EeV).  Results for real data are shown by red
  curves.  Expectations from simulation data sets are blue curves.
  Shaded regions are 95\% containment of results of simulated data
  sets.  Dashed curves are Gaussian approximations for the expected Li-Ma
  distribution.  }
\label{sign-figure}
\end{figure}

Statistical results for the ensemble of celestial targets are shown in
Figure \ref{sign-figure}.  Each row of plots represents one of the
four energy ranges.  Red lines show the distribution of Li-Ma
significance obtained from the data, while blue lines show the
expectation obtained by averaging over simulation data sets.  (Each
simulation data set is analyzed exactly as the real data, using all of
the {\it other} simulation data sets to determine the background for
every target.)  Also shown in each figure is the Gaussian function
that the Li-Ma distribution is expected to approximate if deviations
from expected values are due only to statistical fluctuations.

For each energy range there are two plots.  On the left is the
differential histogram, binned in increments of Li-Ma significance.
On the right are two (unbinned) integral distributions of the same
Li-Ma significances.  One focuses on the tail of high significance by plotting
(for each Li-Ma significance) the total number of targets of equal or greater
significance.  The other focuses on the tail of low significances by
plotting the total number of targets that had equal or lower Li-Ma
significance.  The shaded bands are 95\% containment bands for
simulation data sets.  For any number of targets (plotted vertically),
the shaded band extends horizontally over 95\% of the simulation data
sets; 2.5\% of the simulation integral curves were to the left of the
band at that vertical level, and 2.5\% of the simulation integral
curves were to the right of the band.  

The fact that the red curve does not lie to the right of the right-hand
shaded region means that this search has not identified obviously
significant hot spots.  The deviation from the Gaussian curve for
negative significances in the case $E\geq 3$ EeV is caused by the very low
statistics in many targets.

\subsection{Upper Limits}\label{upper_limits}

\begin{figure}[ht]
\begin{tabular}{cc}
	\includegraphics[width=0.5\textwidth]{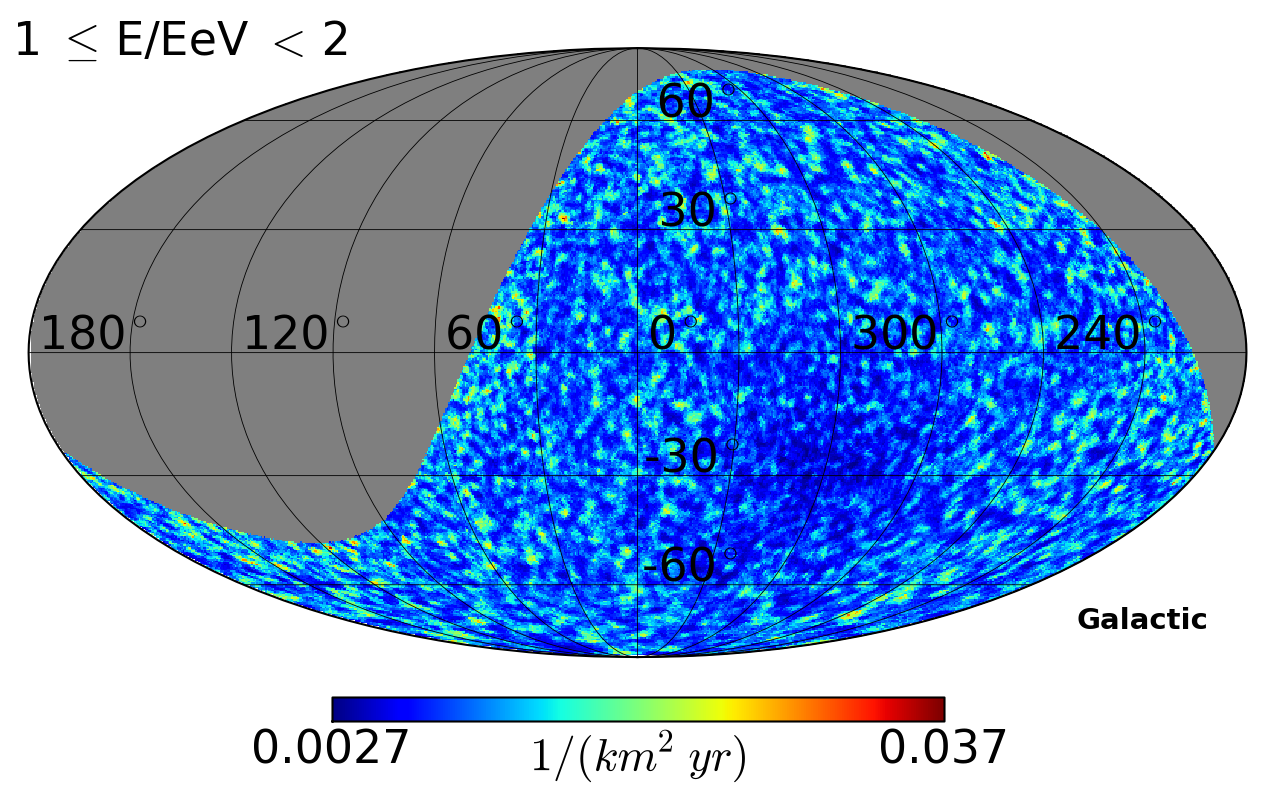} &
	\includegraphics[width=0.5\textwidth]{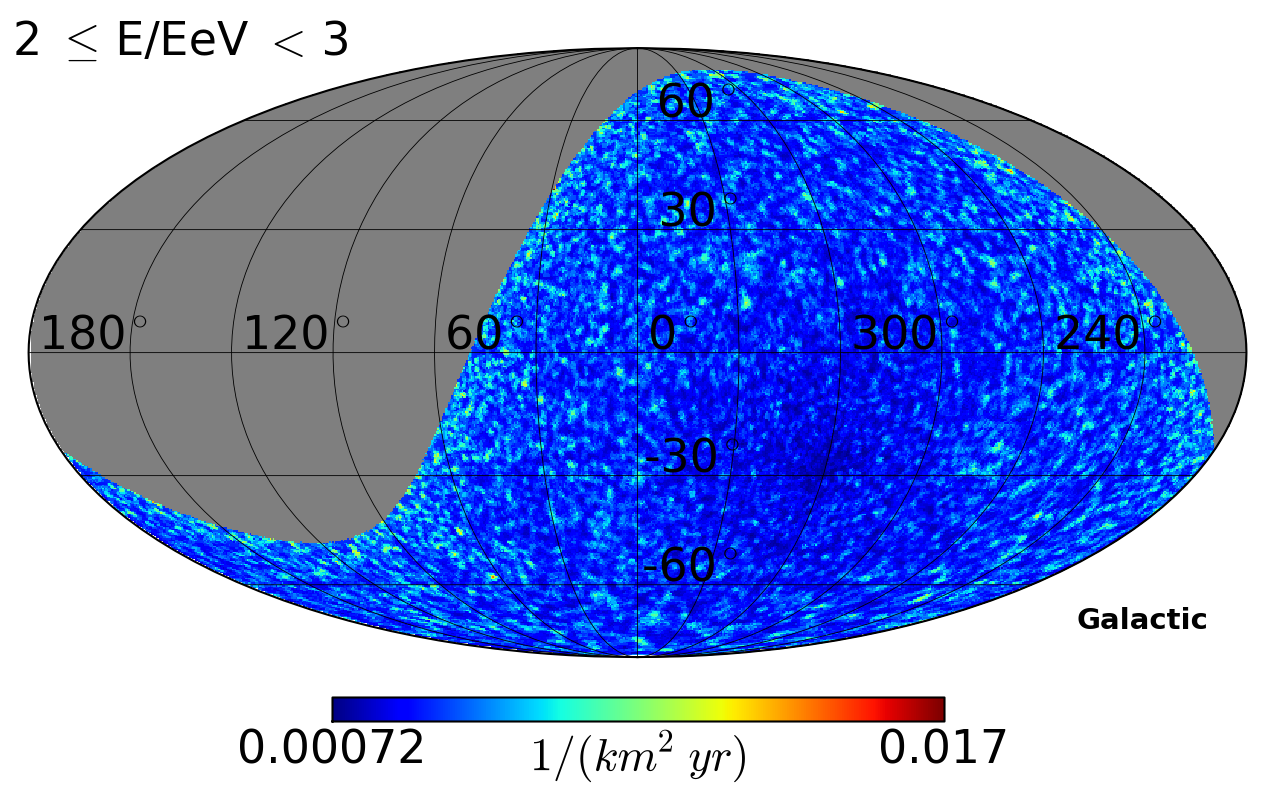} \\
	\includegraphics[width=0.5\textwidth]{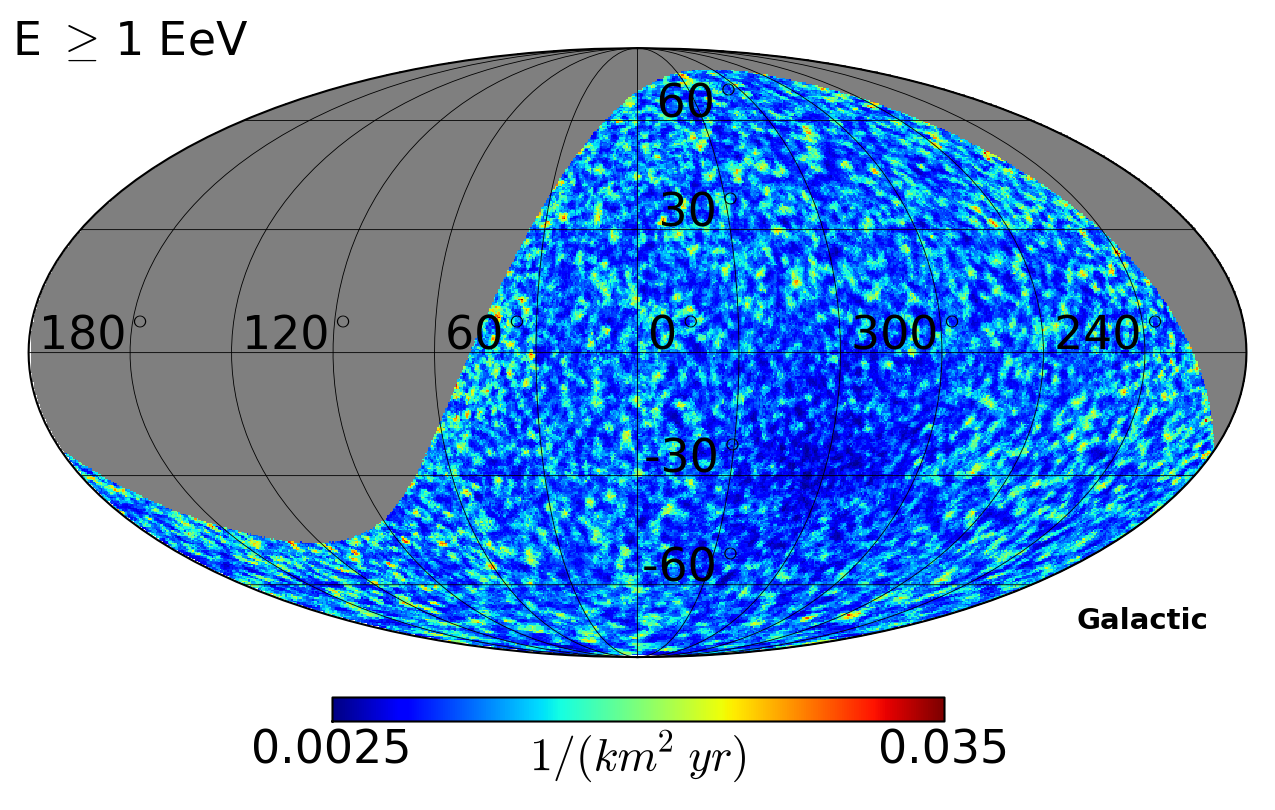} &
	\includegraphics[width=0.5\textwidth]{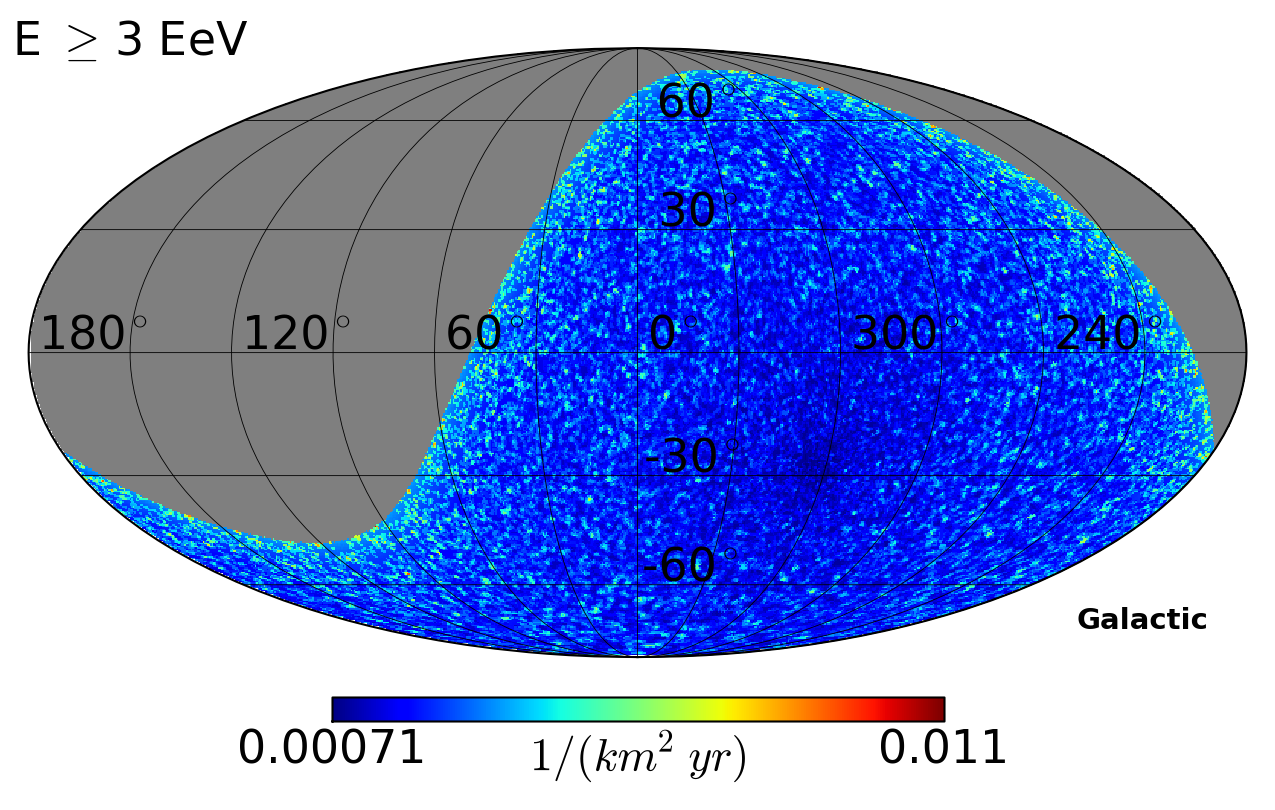} \\
\end{tabular}
\caption{Celestial maps of the flux upper limit
  ($\frac{\mathrm{particles}}{\mathrm{km}^2 \mathrm{yr}}$) in Galactic coordinates.  }
\label{ulimits-figure}
\end{figure}

Flux upper limits (95\% CL) for each target direction are displayed in
the color sky plots of Figure \ref{ulimits-figure}.  Each limit is
calculated according to the method explained in sections
\ref{limit_calc} and \ref{flux-limit}, and it is the upper limit on
the time-averaged neutron flux from that celestial direction.

\begin{figure}[ht]
\includegraphics[width=0.9\textwidth]{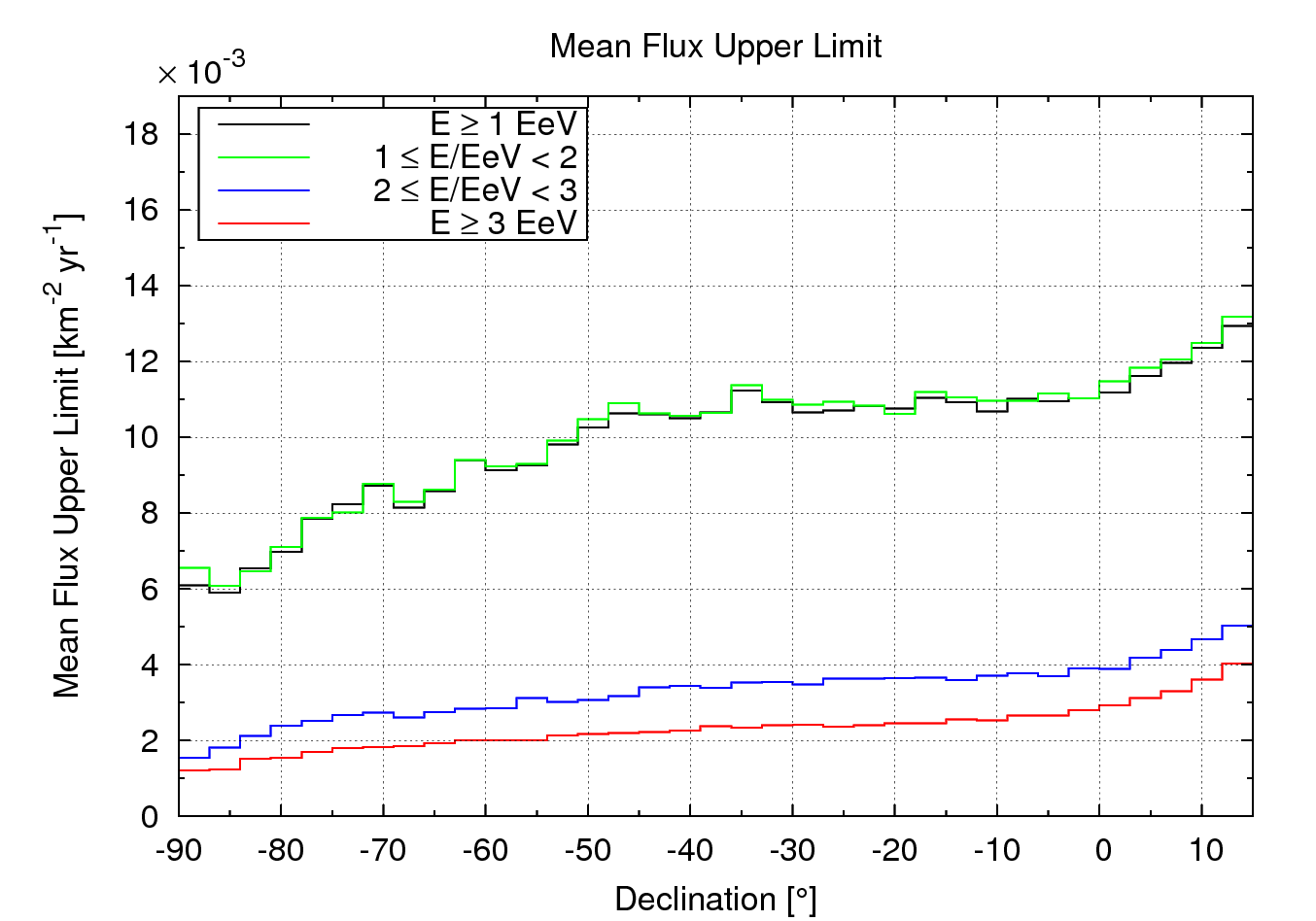} 
\caption{The flux upper limit for each of the four energy ranges,
  averaged over targets in 3-degree bands of declination.}
\label{MeanULs}
\end{figure}

The mean flux upper limit is shown as a function of declination in
Figure \ref{MeanULs} for each of the energy ranges.  The upper limits
tend to be greater (weaker) for the northern declinations where the
directional exposure (shown in Figure \ref{Acceptance}) is reduced.
The limits are lowest (strongest) near the south pole
($-90^\circ$ declination) where the directional exposure is maximum,
but the mean value is less accurately determined in that region
because there are relatively few targets in a declination band.

\section{Summary and discussion}\label{discussion}

The blind search for a flux of neutral particles using the Auger SD
data set finds no candidate point on the sky that stands out among the
large number of trial targets.  Upper limits have been calculated for
all parts of the sky using four different energy ranges.  Three of
those ranges are independent data sets and the fourth is the
combination of the other three.  These upper limits pertain to
neutrons, with systematic uncertainties as discussed in Section
\ref{Uncertainties}.  (The methods used in this paper are less
sensitive to photons.)  

The upper limits are generally more stringent where the directional
exposure is relatively high, but they are strong enough to be of
considerable astrophysical interest in all parts of the exposed sky.
Above 1 EeV, the typical (median) flux upper limit is 0.0114
neutron/km$^2$yr.  That corresponds to an energy flux limit of
0.083 eV/cm$^2$s (or 0.026 EeV/km$^2$yr) in the EeV energy
decade if the differential neutron spectrum is proportional to
$1/E^2$.  Even for the regions of minimum sensitivity, the flux upper
limit does not exceed 0.046 particles/km$^2$yr, corresponding to
0.34 eV/cm$^2$s (or 0.106 EeV/km$^2$yr) for a $1/E^2$
spectrum.

As noted in the Introduction, this energy flux limit is well below
what is observed from some Galactic TeV gamma-ray sources, and
hadronic production of photons by protons with a $1/E^2$ spectrum
should have equal power in each energy decade.  The luminosity emitted
in neutrons should be at least as great as the luminosity emitted in
hadronically produced photons.  The upper limits on neutron fluxes at
EeV energies indicate that TeV gamma-ray emission from those sources
might be of electromagnetic origin or else their proton spectra are
not as hard as $1/E^2$ up to EeV energies.

At EeV energies, there is evidence that the cosmic ray composition
includes a strong proton component \citep{Composition}.  With
reasonable assumptions about the containment time of such protons in
the Galaxy, it can be shown that a neutron flux should be detectable
if the sources of those protons are in the Galaxy and continuously
emitting protons in all directions, assuming the neutron luminosity of a
source is not negligible compared to its proton luminosity and 
there are not more than a few such sources per cubic kiloparsec.  The
{\it absence} of any detectable neutron flux might suggest that the
sources are extragalactic, or transient, or emitting in jets, or
optically thin to escaping protons, or individually weak but densely
distributed.  The flux limits on neutrons couples with the absence of
detectable large-scale anisotropy at EeV energies \citep{Largescale} to
constrain models in which EeV protons are produced by a low density of
strong sources in the Galaxy.

Although no flux of neutrons has been detected in this blind search of
the exposed sky, it is possible that there is a measurable flux of
neutrons from some candidate source of cosmic rays.  There are some
targets with Li-Ma significance greater than $4\sigma$ in this study,
but their number is consistent with what is expected from normal
statistical fluctuations.  The blind search performed here necessarily
involves a very large number of trials.  It is sensible to look
carefully at a small number of astrophysically motivated candidate
source locations.  Results of a targeted search will be reported
separately.

\section*{Acknowledgements}

The successful installation, commissioning, and operation of the
Pierre Auger Observatory would not have been possible without the
strong commitment and effort from the technical and administrative
staff in Malarg\"ue.

We are very grateful to the following agencies and organizations for
financial support: Comisi\'on Nacional de Energ\'ia At\'omica,
Fundaci\'on Antorchas, Gobierno De La Provincia de Mendoza,
Municipalidad de Malarg\"ue, NDM Holdings and Valle Las Le\~nas, in
gratitude for their continuing cooperation over land access,
Argentina; the Australian Research Council; Conselho Nacional de
Desenvolvimento Cient\'ifico e Tecnol\'ogico (CNPq), Financiadora de
Estudos e Projetos (FINEP), Funda\c{c}\~ao de Amparo \`a Pesquisa do
Estado de Rio de Janeiro (FAPERJ), Funda\c{c}\~ao de Amparo \`a
Pesquisa do Estado de S\~ao Paulo (FAPESP), Minist\'erio de
Ci\^{e}ncia e Tecnologia (MCT), Brazil; AVCR AV0Z10100502 and
AV0Z10100522, GAAV KJB100100904, MSMT-CR LA08016, LG11044, LC527,
1M06002, MSM0021620859 and RCPTM - CZ.1.05/2.1.00/03.0058, Czech
Republic; Centre de Calcul IN2P3/CNRS, Centre National de la Recherche
Scientifique (CNRS), Conseil R\'egional Ile-de-France, D\'epartement
Physique Nucl\'eaire et Corpusculaire (PNC-IN2P3/CNRS), D\'epartement
Sciences de l'Univers (SDU-INSU/CNRS), France; Bundesministerium f\"ur
Bildung und Forschung (BMBF), Deutsche Forschungsgemeinschaft (DFG),
Finanzministerium Baden-W\"urttemberg, Helmholtz-Gemeinschaft
Deutscher Forschungszentren (HGF), Ministerium f\"ur Wissenschaft und
Forschung, Nordrhein-Westfalen, Ministerium f\"ur Wissenschaft,
Forschung und Kunst, Baden-W\"urttemberg, Germany; Istituto Nazionale
di Fisica Nucleare (INFN), Ministero dell'Istruzione,
dell'Universit\`a e della Ricerca (MIUR), Italy; Consejo Nacional de
Ciencia y Tecnolog\'ia (CONACYT), Mexico; Ministerie van Onderwijs,
Cultuur en Wetenschap, Nederlandse Organisatie voor Wetenschappelijk
Onderzoek (NWO), Stichting voor Fundamenteel Onderzoek der Materie
(FOM), Netherlands; Ministry of Science and Higher Education, Grant
Nos. N N202 200239 and N N202 207238, Poland; Funda\c{c}\~ao para a
Ci\^{e}ncia e a Tecnologia, Portugal; Ministry for Higher Education,
Science, and Technology, Slovenian Research Agency, Slovenia;
Comunidad de Madrid, Consejer\'ia de Educaci\'on de la Comunidad de
Castilla La Mancha, FEDER funds, Ministerio de Ciencia e Innovaci\'on
and Consolider-Ingenio 2010 (CPAN), Xunta de Galicia, Spain; Science
and Technology Facilities Council, United Kingdom; Department of
Energy, Contract Nos. DE-AC02-07CH11359, DE-FR02-04ER41300, National
Science Foundation, Grant No. 0450696, The Grainger Foundation USA;
ALFA-EC / HELEN, European Union 6th Framework Program, Grant
No. MEIF-CT-2005-025057, European Union 7th Framework Program, Grant
No. PIEF-GA-2008-220240, and UNESCO.


\begin{thebibliography}{}

\bibitem[Aharonian \& Neronov(2005)]{Aharonian}F. Aharonian, A. Neronov, {\it
  Astrophys. J.} {\bf 619}, 306 (2005).


\bibitem[Bellido {\it et al.}(2001)]{Bellido}Bellido, J. A., {\it et al.},
  {\it Astropart. Phys.} {\bf 15}, 167 (2001).

\bibitem[Berezinsky {\it et al.}(2006)]{Berezinsky}V. Berezinsky, G. Gazizov, and S. Grigorieva, {\it
  Phys. Rev.} {\bf D74}, 043005 (2006).

\bibitem[Blumenthal(1970)]{Blumenthal}G.R. Blumenthal, {\it Phys. Rev.} {\bf D1}, 1596 (1970).

\bibitem[Bonifazi(2009)]{AngRes}C. Bonifazi for the Pierre Auger Collaboration, ``The
  angular resolution of the Pierre Auger Observatory,'' {\it
    Proc. 31st ICRC} (Lodz) [arxiv:0901.3138] (2009).

\bibitem[Bossa, Mollerach, \& Roulet(2003)]{Bossa}M. Bossa, S. Mollerach,
  E. Roulet, {\it J. Phys. G: Nucl. Part. Phys.} {\bf 29}, 1409
  (2003).

\bibitem[Crocker {\it et al.}(2005)]{Crocker}R.M. Crocker, M. Fatuzzo,
  J.R. Jokipii, F. Melia, R.R. Volkas, {\it Astrophys. J.} {\bf 622},
  892 (2005).

\bibitem[Rouill\'e-d'Orfeuil(2011)]{icrc2011}B. Rouill\'e-d'Orfeuil
  for the Pierre Auger Collaboration, ``Search for Galactic
  point-sources of EeV neutrons,'' {\it Proc. 32nd ICRC} (Beijing)
  [arXiv:1107.4805] (2011).

\bibitem[Gillessen(2009)]{GCDist}S. Gillessen, {\it Ap. J.} {\bf 692}
  1075. [arXiv:0810.4674] (2009).

\bibitem[Gorski {\it et al.}(2005)]{Healpix}K. M. Gorski, {\it et. al.}, {\it Ap. J.}
  \textbf{622}, 759 (2005).

\bibitem[Hayashida {\it et al.}(1999)]{Hayashida}N. Hayashida {\it et
  al.}, AGASA Collaboration, {\it Astropart. Phys.} {\bf 10}, 303
  (1999).

\bibitem[H.E.S.S.(2011)]{Grillo}H.E.S.S. Collaboration, {\it Astron. \& Astrophys.} {\bf
  528}, A143 [arxiv.1102.0773v2] (2011).

\bibitem[Hillas(1972)]{Hillas1972}A. M. Hillas, {\it Cosmic Rays}, Pergamon Press,
  Oxford (1972).

\bibitem[Hinton \& Hofmann(2009)]{Hinton}J. Hinton and W. Hofmann, {\it Ann. Rev. of
  Astron. and Astrophys.} {\bf 47} (1), 523 (2009).

\bibitem[Li \& Ma(1983)]{Li-Ma} T.-P. Li and Y.-Q. Ma, {\it Ap. J.}
  \textbf{272}, 317 (1983).

\bibitem[Medina Tanco \& Watson(2001)]{Watson}G.A. Medina Tanco, A.A. Watson,
  {\it Proceedings of the 27th ICRC} (Hamburg) p.531 (2001),

\bibitem[Particle Data Group(2010)]{NeutronLifetime} Particle Data Group, {\it J. Phys.} {\bf
  G37}, 075021 (2010).

\bibitem[Pesce(2011)]{Pesce}R. Pesce for the Pierre Auger
  Collaboration, ``Energy calibration of data recorded with the
  surface detectors of the Pierre Auger Observatory: an update'' {\it
    Proc. 32nd ICRC} (Beijing) [arXiv:1107.4809] (2011).

\bibitem[Pierre Auger Collaboration(2004)]{NIM_EA}The Pierre Auger Collaboration, {\it Nucl. Instrum. Meth.},
{\bf A523}, 50 (2004).

\bibitem[Pierre Auger Collaboration(2007)]{GC}The Pierre Auger
  Collaboration, {\it Astropart. Phys.} {\bf 27}, 244 (2007).

\bibitem[Pierre Auger Collaboration(2008)]{Spectrum_PRL}The Pierre Auger Collaboration, {\it Physical
  Review Letters} {\bf 101}, 061101 (2008).

\bibitem[Pierre Auger Collaboration(2010a)]{PLB}The Pierre Auger Collaboration, {\it Physics Letters}
  {\bf B 685}, 239 (2010a).

\bibitem[Pierre Auger Collaboration(2010b)]{Acceptance}The Pierre Auger Collaboration, {\it Nucl.
  Instrum. and Meth.} {\bf A613}, 29 (2010b).

\bibitem[Pierre Auger Collaboration(2010c)]{Composition}The Pierre Auger Collaboration, {\it
  Phys. Rev. Lett.} {\bf 104}, 091101 (2010c).

\bibitem[Pierre Auger Collaboration(2011)]{Largescale}The Pierre Auger Collaboration, {\it
  Astropart. Phys.} {\bf 34}, 627 (2011).

\bibitem[Salamida(2011)]{ICRC_spectrum}F. Salamida for the Pierre Auger Collaboration, 
``Update on the measurement of the CR energy spectrum above $10^{18}$ eV 
made using the Pierre Auger Observatory'' {\it Proc. 32nd ICRC} (Beijing) 
[arXiv:1107.4809] (2011).

\bibitem[Zech(1989)]{Zech} G. Zech, {\it Nucl. Instrum. Methods}, \textbf{A277},
  608 (1989).


\end{thebibliography}
\end{document}